\documentclass[3p]{elsarticle}
\usepackage{lipsum}
\makeatletter
\def\ps@pprintTitle{%
 \let\@oddhead\@empty
 \let\@evenhead\@empty
 \def\@oddfoot{}%
 \let\@evenfoot\@oddfoot}
\makeatother

\usepackage{amsmath,amsthm,verbatim,amssymb,amsfonts,graphicx,natbib,subcaption}

\begin{document}

\begin{frontmatter}

\title{The Structure of the Hadron-Quark Reaction Zone}

\author[cal]{Amir Ouyed\corref{cor2}}
\ead{ahouyedh@ucalgary.ca}

\author[cal]{Rachid Ouyed}
\author[usc]{Prashanth Jaikumar}

\address[cal]{Department of Physics and Astronomy, University of Calgary,
2500 University Drive NW, Calgary, Alberta, T2N 1N4, Canada}
\address[usc]{Department of Physics and Astronomy, California State University Long Beach,
1250 Bellflower Blvd., Long Beach, CA 90840 U.S.A}

\cortext[cor2]{Principal corresponding author}

% Abstract (Do not insert blank lines, i.e. \\) 
\begin{abstract}
Hadron-quark combustion in dense matter is a central topic in the study of phases in compact stars and their high-energy astrophysics. We critically review the literature on hadron-quark combustion, dividing them into a ``first wave'' that treats the problem as a steady-state burning with or without constraints of mechanical equilibrium, and a ``second wave'' which uses numerical techniques to resolve the burning front and solves the underlying Partial Differential Equations for the chemistry of the burning front under less restrictive conditions. We detail the inaccuracies that the second wave amends over the first wave, and highlight crucial differences between various approaches in the second wave. We also include results from time-dependent simulations of the reaction zone that include a hadronic EOS, neutrinos, and self-consistent thermodynamics without using parameterized shortcuts. 
\end{abstract}

% Keywords
\end{frontmatter}

\section{Introduction }

\begin{figure}
    \centering
        \includegraphics[width=0.7\textwidth]{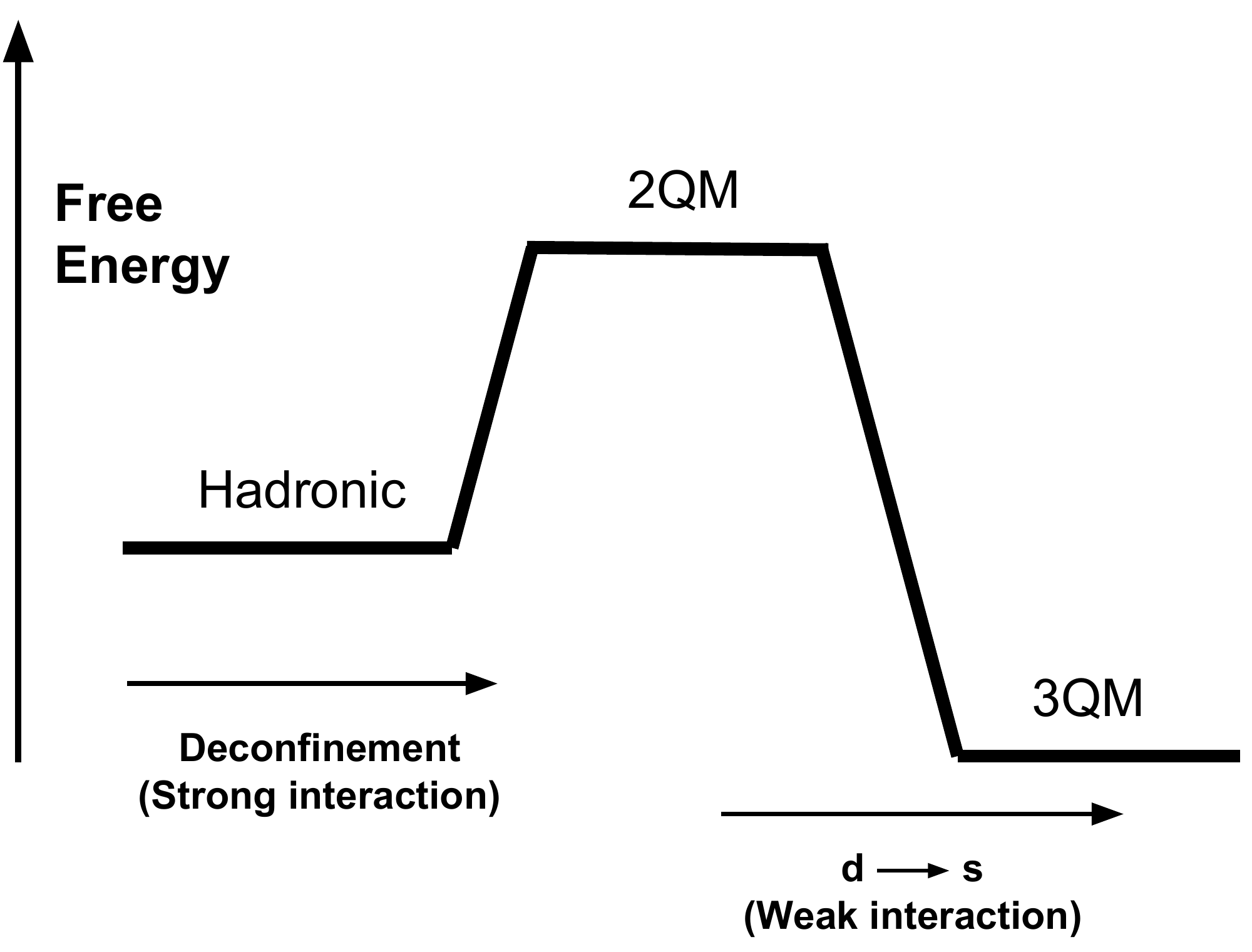}
\caption{Free energy (Helmholtz) diagram of the conversion process of hadronic to 3QM matter.  Hadronic matter deconfines to 2QM through the strong interaction, while 2QM decays to 3QM through the nonleptonic process $d\rightarrow s$ (for the case of deleptonized matter, otherwise there are leptonic processes as well). Notice how the hadronic matter needs to be injected with energy before deconfinement becomes favourable. In the context of a compact object, the high energy is acquired through the high Fermi momenta of the dense hadronic matter. After the hadronic matter is deconfinement into 2QM matter, then the weak interaction can lower its energy to absolutely stable 3QM.}
     \label{chemBarrier}
\end{figure}

Shortly after the hypothesis of absolutely stable strange quark matter (matter made of up, down, and strange postulated) was postulated \cite{bodmer1971collapsed,witten1984cosmic,terazawa1979tokyo}, its astrophysical implications began to be investigated.  It is empirically known that two-flavor quark matter (2QM), matter made of up and down deconfined quarks, is unstable at low temperatures and densities, since otherwise hadrons such as protons and neutrons would decay spontaneously  into 2QM.  However, the hypothesis of absolutely stable strange quark matter (3QM) is that the addition of an extra degree of freedom in the form of strange quarks lowers the free energy per quark of 3QM to the point that it could be more stable than both hadrons and 2QM \cite{bodmer1971collapsed,witten1984cosmic,terazawa1979tokyo}. The fact that we have not seen any 3QM bulk matter in the universe can be reconciled with this hypothesis by realizing that such matter requires that hadronic matter be either heated up or compressed to the point it deconfines into 2QM, and then decays into 3QM. In  Fig. \ref{chemBarrier} we show a free energy diagram on how the transition of hadronic to 3QM would come about.  The horizontal axis represents the direction of the transition (i.e. from hadronic to 2QM and then to 3QM), and the vertical axis represents the free energy. According to the diagram, the system must be injected with sufficient energy, through for example, high fermi-momenta, so that it is energetically favorable for the strong interaction to deconfine hadronic matter into 2QM. Then 2QM decays into 3QM through the weak interaction, mostly through the reaction $d\rightarrow s$, ending up in a lower free energy configuration than before.

Using the basic thermodynamic Bag Model, which treats quark matter as a non-interacting Fermi gas confined by a negative pressure of magnitude $B$, one can show that for certain values of $B$ allowed by the stability of hadronic matter over 2QM, the conversion of hadronic matter to 3QM can release as much as $\sim 100$ MeV per baryon converted \cite{weber2005strange}. Now, if this were scaled up to the conversion of a whole neutron star, which is made of about $10^{57}$ baryons, one would get about $10^{53}$ erg \cite{menezes2006gamma} of binding energy released, about the same order of magnitude as the energy released in a core-collapse supernova.

However, this order of magnitude calculation does not address the dynamic processes through which the conversion appears.  A quantitative understanding of the dynamic aspect of the conversion is necessary in order to find out whether this energy is released in an explosive manner that can be detected by an Earth based observatory. Witten, in his seminal paper \cite{witten1984cosmic}, outlined qualitatively how this time-dependent process would appear. The core of a neutron star may be dense enough to nucleate a 2QM bubble that in turn can decay into 3QM stable matter.  This 3QM core will then accrete the surrounding neutrons (since neutrons are neutral and therefore will not be repelled by the 3QM electric field) and these neutrons will automatically deconfine and merge with the 3QM until the whole neutron star converts into a quark star. The reason why the conversion requires an initial bubble of quark matter to be nucleated is that once there are enough s-quarks in the 3QM matter, hadrons that get absorbed by the 3QM matter can lower their energy by converting as well.

Another way that this conversion could appear is through the creation of a shock, which was first discussed by Benvenuto et al. \cite{benvenuto1989evidence}. If the deconfinement of the quark core appears suddenly, for example in the case of a core-collapse supernova, the sudden increase of density in the iron core could trigger abrupt deconfinement, generating a shock wave that compresses the overlaying matter into 2QM. Then 2QM could decay into 3QM, releasing energy explosively.

How does one study this dynamic combustion process? The high densities and temperatures in the conversion process imply that it is a hydrodynamic problem, since both the hadronic and quark matter would be fluids. Furthermore, since the conversion process turns a substance (hadrons) into another substance (quarks) and this process releases heat, we can model the process as a hydrodynamic combustion process where the interface is formed by ``chemical'' reactions. Furthermore, these reactions will create neutrinos that are transported across the interface. Mathematically, the most accurate way to model this system is to solve a set of Partial Differential Equations (PDEs) coupling reaction rates, diffusion rates, hydrodynamic transport, and neutrino transport. However these PDEs are nonlinear and therefore cannot be solved analytically, and require numerical techniques that are computationally subtle and expensive to implement. Therefore the majority of the studies on the combustion process use semi-analytic, steady state-models, ignoring the time-dependent aspects of the problem (e.g. \cite{baym1985cygnus,olinto1987conversion,olesen1991burning,lugones1994combustion,drago2007burning,furusawa2016hydrodynamical,furusawa2016hydrodynamical2}). Recently, there have been efforts to construct a proper time-dependent, numerical treatment to the combustion process, starting with the seminal work of Niebergal et al. \cite{niebergal2010numerical,niebergal2011hadronic}.  

In these proceedings we will critically review the work done on the hadron-quark combustion process.  We first look at the steady state, semi-analytical treatment of the problem, which was the exclusive method up to the 2010s. We will refer to the bulk of these studies as the ``first wave'' (Section \ref{firstWave}).  Then, we review the studies that took the time-dependent aspect of the combustion process into account, which we term the ``second wave'' (Section \ref{secondWave}). We compare the two and explain which shortcomings were remedied by the second wave (Section \ref{comparison}), and different approaches that are prevalent in this ``second wave''. Finally we will conclude with a statement about future work (Section \ref{conclusion}).

\section{ First Wave}\label{firstWave}

There are roughly two kinds of approaches to the the steady state viewpoint. The first is the semi-analytic method that treats the system in mechanical equilibrium as well as steady state, which originated with Olinto's approach \cite{olinto1987conversion}. The other treatment uses hydrodynamic jump conditions, which are under steady state assumptions, but the interface is not in mechanical equilibrium, and there is a discontinuity across the interface with thermodynamic variables such as density and pressure. We may call this as the jump-condition approach. 

Olinto's approach \cite{olinto1987conversion} assumes that the reaction zone is in steady state and also in mechanical equilibrium, with no shock-like discontinuities.  Furthermore, the strange quark number density needed to make the conversion favorable was parameterized, not deduced self-consistently from the equations of state.  It was found that the speed of the conversion front was up to 20 km/s \cite{olinto1987conversion}. Olinto's solution to the burning speed has a neat analytic expression:

\begin{equation}\label{olintoEq}
         v=\sqrt{\frac{D}{\tau}\frac{(1-{Y_s}_0)^4}{1-{Y_s}_0^4}}\,,
     \end{equation}
     where ${Y_s}_0$ is the minimum strange quark fraction required for stable strange quark matter, $\tau=10^{-8}$s is the timescale of the nonleptonic interaction, $D$ is the diffusion coefficient.  This formula has been used by many studies (e.g. \cite{drago2015combustion})

The jump-condition approach was originally used by Benvenuto et al. \cite{benvenuto1989evidence} to model a detonation shock that compresses the overlaying matter and converts it into quark matter. However, the jump-condition wasn't just relegated to the treatment of shocks, but was used for modelling deflagration interfaces that are subsonic (\cite{lugones1994combustion},\cite{drago2015combustion}).  In the latter case, the assumption made is that the reaction zone is small enough to treat it as a discontinuity, and that the dynamics of the reaction zone cam be safely ignored.  Another frequently used approximation are the polytrope-like EOS where the temperature is ignored and the energy density is simply a function of density (\cite{olesen1991burning},\cite{lugones1994combustion}).

 Since most studies do not resolve the reaction zone and simply treat it as a discontinuity, a ``rule of thumb'' must be constructed that indicates when it is plausible for the hadronic fuel to convert into quark ash. In the realistic scenario, the conversion of hadronic to quark matter must appear if it is energetically favourable, in other words if it decreases the free energy of the system.  However in order to track the free energy of the reaction zone, the resolution of the study must be fine enough to take into account weak interaction terms and heat transfer across the reaction zone \cite{furusawa2016hydrodynamical}. Given that most studies approximate the reaction zone as a discontinuity and consequently do not track this free energy, an approximation must be used. One of the more popular approximations is Coll's Condition \cite{coll1976combustion} which can be written  as:
 
\begin{equation}\label{inequality}
    e_h(P,X)>e_q(P,X)\,,
\end{equation}
where $e_h$ stands for the energy density of hadronic fuel, and $e_q$ for the energy density of quark matter, $P$ for pressure, and $X$ for dynamical volume.   
In this approximation, the hadronic fuel in a given zone changes to quark ash if the energy density of the fuel is higher than the energy density of the ash for the same pressure and same dynamical volume.
 Studies have shown that the above condition does not lead to the complete combustion of a neutron star to a quark star, and instead leads to a hybrid star with a quark core \cite{herzog2011three}.

\section{Coll's condition}
Coll's condition has an important role in the history of  hadron-quark combustion research. It was used for the first time by Lugones et al. \cite{lugones1994combustion}, and they were the first to point out that this condition could quench combustion at lower densities, or not allowed at all, depending on the stiffness of the quark EOS.  The next major use was with Herzog's et al. \cite{herzog2011three} where they studied numerically in three dimensions the combustion of a neutron star to a quark star. However given that the resolution was too coarse to simulate the reaction zone, the laminar burning speed was parameterized and Coll's condition was imposed as a heuristic that dictated whether combustion is allowed or not at each computational zone. Drago et al. \cite{drago2015combustion} explored Coll's condition more fully, and  argued that if Coll's condition is satisfied, the combustion becomes turbulent while if it is not satisfied, the combustion becomes diffusive, laminar and slower. 

However, we agree with Furusawa et al. \cite{furusawa2016hydrodynamical} that Coll's condition is not justified under rigorous physical and thermodynamic grounds, and instead is postulated as an apriori hypothesis \cite{coll1976combustion}. This is because the question of the possibility of combustion can be solved with thermodynamics, without the use of some aprioristic condition like Coll's: combustion simply appears when the conversion between hadronic to quark matter is energetically favourable, which is a question of comparing the helmholtz free energy per baryon of both the hadronic state and the quark state, without having to appeal to the internal energy at a specific pressure and dynamical volume (where the pressure and dynamical volume are assumed to be equal at both sides, since both sides of the inequality \eqref{inequality} share the same pressure $P$). Since the strong interaction is responsible for deconfinement, the timescale of deconfinement is $\tau_{d}\sim 10^{-23}$s. If we assume the width of the interface is at most $l\sim cm$ \cite{niebergal2011hadronic}, and choose the speed of light $\sim c$ as the sound speed, then the sonic time is about $t_{s}\sim l/c \sim 10^{-11}$s.  Given that even for the case where the sound speed is close to the speed of light, $\tau_{d}\ll \tau{s}$, one cannot assume sound waves smooth out the pressure gradient to the point where pressure $P$ is equal to both sides of the interface. Therefore both the use of internal energies instead of free energies, and the assumption of pressure being equal across the interface are not assumptions backed with physical arguments.  We must conclude then, in the case of the past work of Lugones et al. \cite{lugones1994combustion} and Herzog et al. \cite{coll1976combustion}, the use of this condition is not justified. 

Recently, Drago et al. \cite{drago2015combustion}  have given a different interpretation to Coll's condition. They argued that the failure of Coll's condition does not prevent hydrodynamic combustion, but simply suppresses turbulence, and which makes combustion proceed in a diffusive and slower way. This is because a satisfaction of Coll's condition implies that the energy density of the quark ash is lower for the same pressure $P$ as the energy density of the hadronic ash, which triggers Rayleigh Taylor instabilities and therefore turbulence.  This interpretation of Coll's condition, although more physically informed, still has some problems. The first issue is that the condition uses the same pressure $P$ across both sides of the interface. As mentioned before, deconfinement acts in a much faster way than the sound waves, so once conversion comes into effect, the quark EOS will abruptly soften, giving the quark side a lower pressure than the hadronic side immediately after deconfinement (since sound waves are comparatively slow). Therefore it does not follow that the quark ash  will necessarily have a higher energy density if Coll's condition is violated. The second issue is that Drago et al. use Olinto's equation \eqref{olintoEq} to calculate the burning speed once Coll's condition is violated. The issue here is that Eq. \eqref{olintoEq} is derived under the assumption of mechanical equilibrium, which may lead to much slower burning speeds than in the realist case, where mechanical equilibrium is not assumed apriori.  Although Drago et al's interpretation is definitely more plausible and physically informed than others uses of Coll's condition, arguing that the failure of Coll's condition suppresses turbulence is a very strong conclusion in light of the known physical processes.

\section{Second Wave Studies }\label{secondWave}

The first time-dependent studies of the hadron-quark reaction zone appeared in 2010 with the seminal work of Niebergal et al. \cite{niebergal2010numerical}. They focused on the reaction zone itself and treated it as a dynamical system where neutrino physics, hydrodynamics, weak processes, and diffusion processes were coupled. They solved the partial differential equations (PDEs) using numerical techniques.  However, they approximated the hadronic fuel as unburned two flavored (2QM) quark matter, and modelled the neutrinos with a leakage scheme. One of their important results is a burning speed of $0.001$c - $0.01$c which is six orders of magnitude higher than typical speeds. They also parameterized neutrino cooling as a temperature diminution across the jump conditions.  They found that for a given amount of cooling, the interface could halt. This hinted that there are instabilities caused by the interaction of neutrinos and hydrodynamics. 

Later on, Ouyed et al.\cite{ouyed2018numerical} explored the effect of neutrinos in the interface by implementing a neutrino diffusion scheme and including neutrinos and electrons in the EOS.  They found that indeed, neutrinos can cause the interface to halt.  This is because pressure is lowered behind the interface by the loss of neutrinos and an electron pressure gradient opposes the motion of the interface. These studies not only confirmed the previous semi-analytic studies on neutrino physics done by Niebergal et al. but also discovered that the effects of neutrinos and electrons are, at the very least, as important as the effects of the high density EOS. Recently Ouyed \cite{ouyed2019neutrino} extended the study by using a hadronic EOS, and discovered a thermal instability that quenches the burning, since the heating of hadronic matter by neutrinos raises a free energy barrier that prevents the burning. 

In  Fig. \ref{interfaceBarrier} we show two snapshots of the interface: in the Upper Panel neutrino absorption by matter is turned off, while in the Lower Panel, neutrino absorption is turned on. We can see how the turning on of absorption generates a divergence in the free energy of hadronic and quark matter across the interface: since neutrino absorption lowers the free energy of hadronic matter and also increases the free energy of quark matter, quark matter's free energy becomes too high to allow the conversion of hadronic to quark matter.

In contrast to studies involving Coll's Condition, they resolved the reaction zone sufficiently to track the free energy across the interface to calculate self-consistently when combustion is allowed (Fig \ref{interfaceBarrier}).

In other words, including the dynamical effects of the reaction zone can lead to counter-intuitive behavior such as quenching due to time-dependent effects, which are not captured when using the steady state solutions of Olinto or the jump-conditions \cite{ouyed2018numerical,ouyed2019neutrino}. 

\section{Comparison between first-wave and second-wave}\label{comparison}

After Niebergal's work in 2010 \cite{niebergal2010numerical}, the inclusion of time-dependent effects has led to radically different solutions than the ones of steady-state.  The main differences are:
\begin{enumerate}
    \item Burning speeds that are up to six orders of magnitude faster (0.01 c) than other approaches. For example Olinto's solution gives a lower bound of about $~100$ cm/s.
    \item Quenching due to neutrino leakage under-pressuring the quark ash.
    \item Quenching due to heating of the hadronic fuel by neutrinos, which generates a free energy barrier. 
\end{enumerate}

\begin{figure}
    \centering
        \includegraphics[width=0.7\textwidth]{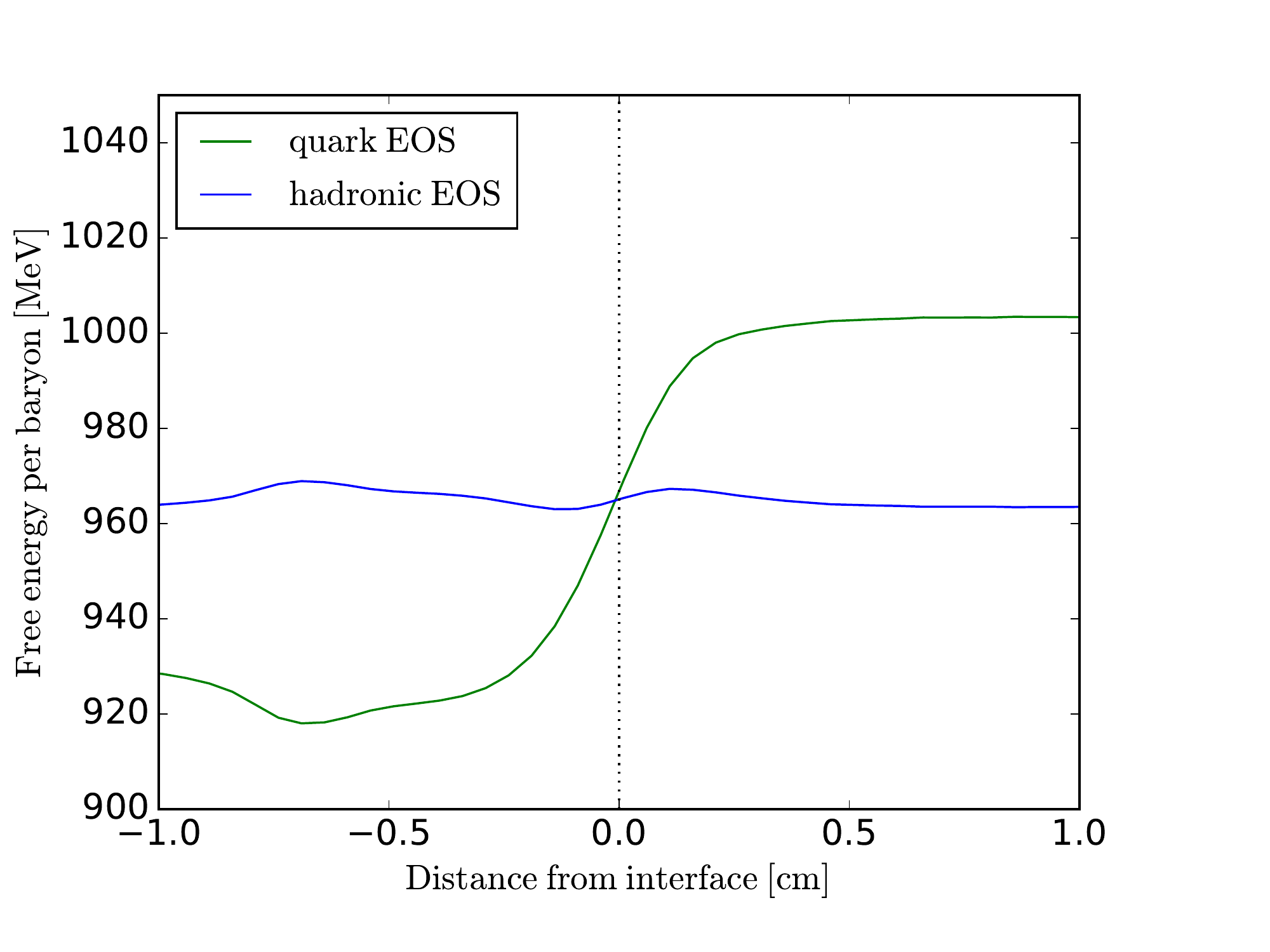}
      \includegraphics[width=0.7\textwidth]{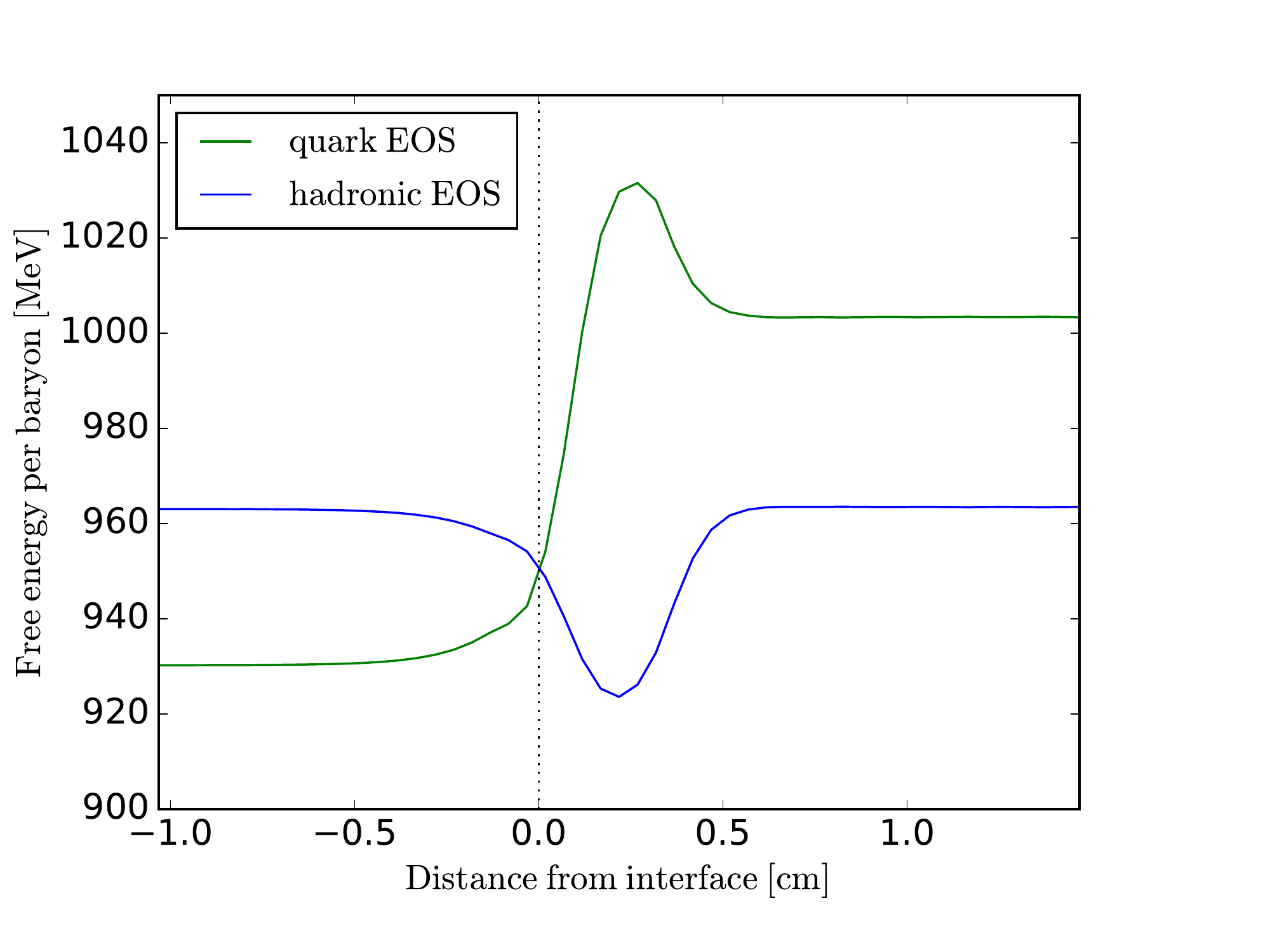}

      \caption{Helmholtz free energy per baryon versus distance from interface. Upper panel depicts the interface with neutrino absorption turned off, while in the Lower Panel, neutrino absorption is turned on. The initial hadronic density was $n_B=0.2$ fm $^{-3}$. We can clearly see that the free energies of quark matter is lower in the left side of the interface than hadronic matter. This implies that a quark matter ash exists in the left side since it is energetically favourable. Furthermore, in the Lower Panel, we can clearly see a free energy barrier forming, with the free energies of quark matter and hadronic matter diverging in the right side of the interface. }

     \label{interfaceBarrier}
\end{figure}

These important differences arise from the fact that the steady-state approximations employed by previous literature are simply too restrictive. Furthermore the collapse of the reaction zone to a discontinuity assumes that the reaction zone is dynamically unimportant, but this is a flawed or over-simplified assumption.  The main inaccuracies in the older approximations are the following:

\begin{enumerate}
\item  For Olinto's approach, the interface is assumed to be in mechanical equilibrium, whereas pressure gradients in the interface can considerably modify the mechanics of the reaction zone.
\item The Navier-Stokes equations has a nonlinear term $v\nabla v$ where $v$ is velocity that is ignored when using Olinto's approach or the jump conditions. Recent studies show that this nonlinear term can lead to quenching behavior such as the one caused by under pressurization due to neutrino leakage.  This effect is virtually ignored by all studies except for some second wave studies \cite{ouyed2019neutrino,ouyed2018numerical,niebergal2010numerical}. 
\item  Coll's condition imposes a very strong constraint for burning, since it prohibits combustion at densities at the same order of magnitude as nuclear saturation density. However as shown by \cite{ouyed2019neutrino}, once one resolves and simulates the reaction zone with its time-dependent dynamics (Fig. \ref{interfaceBarrier}), there is no need for Coll's condition, since the second law of thermodynamics allows combustion for all fuel densities -  all we need to assume is that the hypothesis of absolutely stable quark matter is true. Furthermore, Coll's condition is merely a hypothesis assumed apriori (see \cite{furusawa2016hydrodynamical}) with little physical justification. Combustion is allowed (or not) depending on the behavior of the free energy, not the energy density. Furthermore, Coll's condition assumes the fuel and the ash are in mechanical equilibrium which is not true. 
\item  Not all recent studies of hadron-quark combustion track entropy generation.  Weak processes are intimately coupled to entropy generation since the change of quark and lepton composition impose an entropy source/sink. This profoundly affects the dynamics of the flame, since the EOSs have factors that depend on temperature, and a higher temperature means a stiffer EOS. It follows that the change in entropy can impose pressure gradients into the interface, affecting the speed of the front. 
\item  The commonly used expression for the width of the burning front is inaccurate. The usual formula is taken from Landau et al. \cite{landau2013fluid} is $l=\sqrt{D \times \tau}$ where $D$ is the diffussion coefficient and $\tau$ is the timescale of the weak interaction.  However this formula is based on diffusion only. In a realistic scenario, hydrodynamic processes widen the flame, since hydrodynamic timescales are faster than diffusion timescales in the case of quark matter. This leads to a flame that is six orders of magnitude wider. 
\item Some studies use a zero temperature EOS. In the case of the quark EOS, this would lead to inaccurate results because the front velocity is very sensitive to the temperature of the quark ash. (see \cite{niebergal2010numerical}). Furthermore, when used in conjunction with the jump-conditions, this approximation becomes inconsistent, since a shock is an entropy producing process. 
\end{enumerate}

An important question that arises from the quenching due to neutrino absorption depicted in Fig. \ref{interfaceBarrier} is whether combustion can proceed since neutrino absorption introduces an energy barrier.  However, one has to keep in mind that the neutrino absorption used here is done under the approximation of trapped neutrinos \cite{ouyed2018numerical}. In reality, matter is only semi-opaque in the reaction zone to neutrinos, since the mean free path is of the same order of magnitude as the length of the reaction zone \cite{ouyed2018numerical}.  In reality, the real effect of neutrinos on quenching would be in the middle of the two cases depicted in Fig. \ref{interfaceBarrier} (free streaming versus neutrino absorption) \cite{ouyed2019neutrino}. Furthermore, the energy barrier is very dependent to the specifics of the EOS \cite{ouyed2019neutrino}, so it could be that a different configuration of EOSs wouldn't lead to that free energy barrier. Another important aspect of the barrier is that it is due to the temperature raising by neutrino absorption \cite{ouyed2019neutrino}, so cooling would remove the  energy barrier, restarting the combustion process.

Finally, we showed in \cite{ouyed2019neutrino} that neutrino physics would trigger instabilities very similar to Rayleigh Taylor, since deleptonization across the interface would lead to a pressure gradient that  pushes against a stratified liquid, which is analogous to gravity pushing against a stratified liquid. So it is very probable that this turbulence would kick-start the combustion, although more sophisticated, multidimensional studies are required to explore this issue.

Now, we must mention some words on the limitations of "Second Wave" studies.  The reaction zone has a length scale of $0.1$ cm to $1$ cm. Yet, the whole compact star has a length scale of $10$ km.  Second wave studies resolve the reaction zone at the expense of simulating the whole compact star, since given the disparate scales, the resolution required to simulate the reaction zone is too high to make the computation of the whole compact star feasible. However,  there are large-scale processes and feedback loops that may affect the structure of the reaction zone itself but may require the simulation of the whole compact star. For example, gravity can lead to Rayleigh-Taylor instabilities, and these instabilities require the resolving of the whole compact star.

Herzog et al. \cite{herzog2011three} tried to take the ``best of both worlds'' approach by parameterizing Niebergal's results \cite{niebergal2010numerical} into a function that calculates the laminar velocity of the interface, and using that function in a code  that resolves the global scale  of the compact star. The interface is treated as a discontinuity and all the information of the reaction zone is contained in the parameterized function. Although the resolution of global scales shines light unto important aspects of the combustion problem, such as the Rayleigh Taylor instabilities, it suffers from some limitations. For example, strong non-linearities that permeate the reaction zone's structure, such as neutrino physics,  can cause the front to accelerate, quench, or become unstable \cite{ouyed2018numerical,ouyed2019neutrino}, are effectively suppressed by the approximation of Herzog et al.. Furthermore, Herzog et al. are forced to use Coll's condition in order to trigger combustion, which enforces a very strong constraint where the combustion is quenched before the whole compact star is converted, leaving kilometers of unburnt hadronic matter above the quark core. This constraint would lead to a much weaker electro-magnetic and neutrino signal \cite{ouyed2019neutrino}.

A solution to this problem, namely the extremely different scales of the reaction zone and the compact star, would require much more sophisticated computational techniques than the ones already used. For example, there would have to be adaptive mesh refinement techniques were the regions outside the reaction zone would have to have much coarser resolution than the interface itself.

\section{Conclusion}\label{conclusion}

In conclusion,  in the case of the hadron-quark combustion case, numerical simulations that solve the partial differential equations directly yield very different results from semi-analytic calculations, with front speeds that are several orders of magnitude higher for the former.  We point out that even if numerical simulations solve the time-dependent Navier-Stokes equations (e.g. \cite{herzog2011three}), resolving the interface is necessary, since approximations like Coll's condition may quench the burning, while properly resolving the flame can make combustion always thermodynamically favorable if the hypothesis of absolutely stable strange quark matter is true.

\section*{Acknowledgements}
R.O. is funded by the Natural Sciences and Engineering Research Council of Canada under Grant No. RT731073. P.J. is supported by the U.S. National Science Foundation (NSF) under Grant No. PHY1608959.

\section*{Author Contributions}

A.O. wrote the initial draft of the paper and created the figures and performed the calculations. R.O and P.J edited, provided input for the physical details and interpretations, and approved of the final version of the publication. 
\bibliography{Draft2Bib}

\bibliographystyle{apalike}

%%%%%%%%%%%%%%%%%%%%%%%%%%%%%%%%%%%%%%%%%%

%%%%%%%%%%%%%%%%%%%%%%%%%%%%%%%%%%%%%%%%%%
\end{document}